\documentclass[]{spie}  

\usepackage{amsmath,amsfonts,amssymb}
\usepackage{graphicx}
\usepackage[colorlinks=true, allcolors=blue]{hyperref}

\title{Atmosphyre: Modelling Atmospheric Chromatic Dispersion for Multi-Object Spectrographs}

\author[1*]{Jay Stephan}
\author[1]{Ruben Sánchez-Janssen}
\affil[]{UK Astronomy Technology Centre, Blackford Hill, Edinburgh EH9 3HJ, United Kingdom}

\authorinfo{*E-mail: Jay.Stephan@STFC.ac.uk, github: https://github.com/JamianStephan/atmosphyre
}

\begin{document} 
\maketitle

\begin{abstract}
 The wavelength dependent refraction of light in the atmosphere causes the chromatic dispersion of a target on the focal plane of an instrument. This is known as atmospheric dispersion, with one of the consequences being wavelength dependent flux losses which are difficult to minimise, requiring analysis in both instrument design and operations. We present Atmosphyre, a novel python package developed to characterise the impact of atmospheric dispersion on a spectrograph, with a focus on fibre multi-object spectrographs (MOS) which will be at the forefront of ground-based astronomy for the next few decades. 

We show example simulations and provide recommendations for minimising fibre MOS flux losses. We conclude that the guiding wavelength should typically be bluer than the observing band mid-wavelength, around 25-45\% of the way through the band. The aperture should be centred on this wavelength's location on the focal plane. This wavelength/position remains constant for all reasonable declinations and target hour angles. We also present an application of the package to MOSAIC, the ELT’s multi-object spectrograph. We find that differential losses greater than 10\% are unavoidable for 1h observations that are a) after a local hour angle of 2.5h, or b) at declinations below -60\textdegree \ and above 10\textdegree. We identify that the introduction of an atmospheric dispersion corrector (ADC) would result in the significant reduction of spectral distortions, a gain in survey speed for many observations, and enable the implementation of wider visible observing bands; as a result, there has been a proposal to adopt ADCs at a positioner level for MOSAIC. Future work includes adding field differential refraction to Atmosphyre, important for future wide-field multi-object spectrograph projects such as the proposed WST.

\end{abstract}


\section{INTRODUCTION}
\label{sec:introduction}  

The impacts of atmospheric refraction on ground-based astronomical instrumentation have been well studied over the last few decades \cite{Fillipenko, Donnely, Cuby, Sanchez}. One of the key effects is the chromatic dispersion of a target object due to the wavelength dependence of the refractive index of air - this is referred to as atmospheric dispersion. The dispersion for a wavelength relative to another is given by\cite{Smart}:

\begin{equation}
\Delta R = 206265 \cdot [ n(\lambda) - n(\lambda_{ref})] \cdot tan(z)
\label{eq:dispersion}
\end{equation}
where $\Delta R$ is the displacement in arcseconds, $n(\lambda)$ is the wavelength-dependent atmospheric refractive index, and $z$ is the zenith angle of the target object.

One of the consequences for a spectrograph is wavelength dependent losses as the object has been chromatically dispersed around an aperture, lowering the fraction of a given wavelength's flux entering that aperture. These losses result in a) a lower signal-to-noise for a given exposure time, and b) the introduction of a spectral distortion. This in turn results in longer observation times needed for some proposals and can present a risk to science cases where the spectral shape needs to be well preserved. Due to this, losses must be minimised during an instrument's operations and design.

In operations, losses are reduced (to a limit) through observing constraints and the careful positioning of the aperture\cite{Donnely, Sanchez}. This includes the precise placing of fibres, and the careful alignment of slits relative to the parallactic angle (the axis on which atmospheric dispersion occurs). Even with these considerations, losses can be significant; for example the VLT's VIMOS suffered around 15-20\% losses at the extremes of a spectrum during typical observations \cite{Cuby, Sanchez}. In design, the aperture size and the observing wavelength bands are key deciders on the impact of atmospheric dispersion. In the latter case, as losses are more severe for bluer wavelengths due to stronger refraction\cite{Cuby, Wehbe}, it can be desirable to have narrow visible bands - contrary to what is needed in a modern work-horse spectrograph. A solution is the use of atmospheric dispersion correctors (ADCs), however they are costly and difficult to implement over wide field of views and wide spectral bandwidths\cite{Cuby}. It is straightforward to conclude that the problems posed by atmospheric dispersion are challenging, especially for wide-field multi-object spectrographs (MOS).

There are many wide-field MOS currently under development, representing a massive expansion of ground-based spectroscopic capabilities and placing multi-object spectrography at the forefront of astronomy. This includes instruments recently delivered or nearing delivery for current-generation telescopes such as the VLT's MOONS\cite{MOONS}, VISTA's 4MOST\cite{4MOST}, and WHT's WEAVE\cite{WEAVE}; those for next-generation telescopes such as the ELT's MOSAIC\cite{Hammer} and the GMT's GMACS\cite{GMACS}; and the proposed WST facility\cite{WST}. It is important to characterise atmospheric dispersion effects for these instruments and consider them in in design and operations. However, each time the analysis typically starts from scratch.

This paper presents the early development and release of Atmosphyre, a python package made for the straightforward characterisation of atmospheric dispersion impacts on spectrographs, with a focus on fibre MOS. In Section \ref{sec:simulations} of this paper we present the key steps of the package's simulations and modelling. In Section \ref{sec:results} we present results from Atmosphyre, including an analysis for the ELT's MOSAIC that was carried out during an ADC trade-off study. In Section \ref{sec:conclusions}, we present a summary of this work alongside the next steps in the package's development.

\section{Simulations}
\label{sec:simulations}

The primary inputs for our simulations is the instrument configuration (wavelength range, telescope diameter, and aperture) and the observation details (observing latitude, target declination, target local hour angles, and environmental conditions). At any snapshot during the observation, there are two main stages in the atmospheric dispersion simulation:

\begin{enumerate}
  \item The point-spread functions (PSFs) for the observing wavelengths at that airmass must be modelled - see Section \ref{sec:PSFmodelling}.
  \item    The displacements of the PSFs relative to the aperture and their subsequent flux injection efficiency into the aperture must be calculated - see Section \ref{sec:Dispersionandflux}.
\end{enumerate}

For that snapshot, this provides an injection efficiency curve which describes what fraction of flux enters the aperture as a function of wavelength. By repeating this for many regularly-spaced snapshots in the observation, we can take the time-average to generate the instrument's total injection efficiency curve for that observation.

\subsection{PSF Modelling}
\label{sec:PSFmodelling}

\begin{figure}[h]
\includegraphics[scale=0.8]{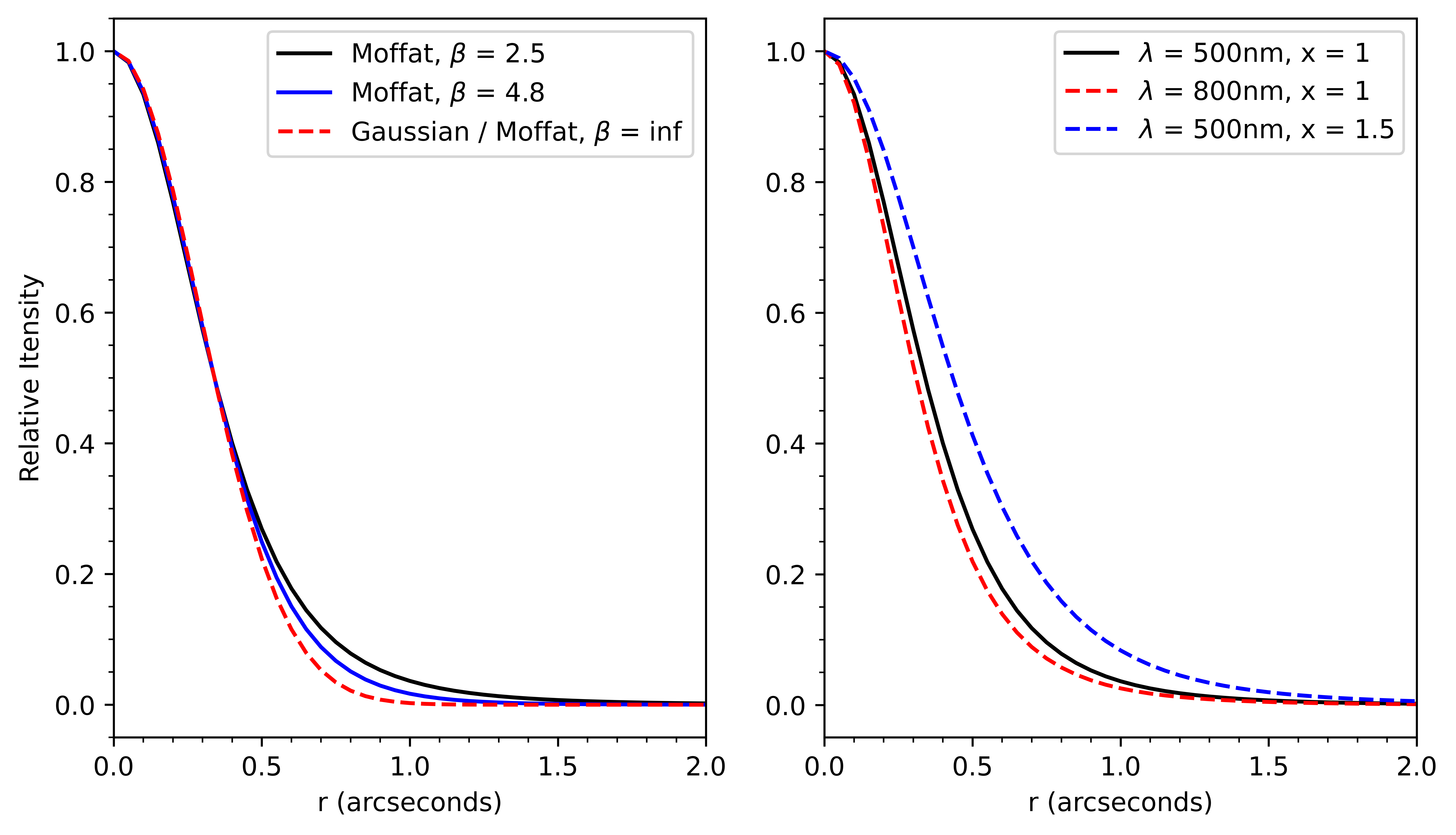}
\centering
\caption{Left: Moffat intensity profiles with $ FWHM = 0.68 $" and various $\beta$ values. $\beta = 2.5$ represents observed PSFs well, $\beta = 4.8$ is the theoretical atmospheric PSF from atmospheric turbulence theory, and $\beta = \infty$ produces a Gaussian profile\cite{Trujillo}. The flux present in the wings grows with smaller values of $\beta$; it is vital to include this effect in injection efficiency simulations for accurate results. Right: Moffat intensity profiles with $\beta = 2.5$ for differing wavelengths and airmasses. $X$ = 1.5 corresponds to a zenith angle of roughly 46\textdegree. We see a broadening of the PSF with shorter wavelengths and larger airmasses.}
\label{fig:moffats} 
\end{figure}

Telescope and atmospheric effects introduce significant wings to a PSF which are important when calculating flux injection efficiency for PSFs not well-centred with respect to an aperture. In this work, the PSF of a monochromatic wavelength on the instrument focal plane is modelled as a Moffat profile\cite{Trujillo}:

\begin{equation}
    PSF(x,y) = \frac{\beta - 1}{\pi \alpha^2} \cdot \left[ 1 + (\dfrac{x^2 + y^2}{\alpha^2})\right]^{-\beta}
\end{equation}
Here, $x$ and $y$ are positions away from the PSF centre; $\beta$ is a constant representing the prominence of the PSF wings ($\beta = 2.5$ is known to provide a good parametrisation for observed PSFs\cite{Trujillo, CASU} and is therefore set as the default value in Atmosphyre, see Fig. \ref{fig:moffats}); and $\alpha$ is a seeing related constant given by:

\begin{equation}
\alpha= FWHM / (2 \sqrt{2^{1/\beta}-1})
\end{equation}
Here, $FWHM$ is the full width at half-maximum of the PSF on the focal plane. The modelled FWHM is calculated as\cite{EsoETC}:

\begin{equation}
    FWHM_{IQ} = \sqrt{FWHM_{tel}^2+FWHM_{atm}^2}
\end{equation}
where $ FWHM_{tel} $ and $FWHM_{atm}$ are the telescope and atmospheric contributions added in quadrature. The former is simply the diffraction limited PSF FWHM, $ \lambda / D$, where $\lambda$ is the wavelength and $D$ is the telescope diameter in the same units. The atmospheric term is given by \cite{EsoETC}:
\begin{equation}
    FWHM_{atm} = s \cdot X^{0.6} \cdot \left(\dfrac{\lambda}{500}\right)^{-0.2} \cdot \gamma
\end{equation}
where $s$ is the seeing measured at 500nm at zenith in arcseconds, $X$ is the airmass, $\lambda$ is the wavelength in nm, and $\gamma$ is a constant dependent on the Fried parameter, the wave-front outer-scale, and the telescope diameter – see Appendix \ref{sec:atmosphericFWHM} for more information. 

The atmospheric term dominates at visible and near-infrared wavelengths, meaning that the PSF broadens with shorter wavelengths and larger airmasses - see Fig. \ref{fig:moffats}.

\subsection{Dispersion and Flux Injection}
\label{sec:Dispersionandflux}

The movement of light on the focal plane due to atmospheric dispersion can be described in three parts:

\begin{enumerate}
\item Dispersion occurs relative to the telescope guiding wavelength as this wavelength will remain fixed on the focal plane throughout an observation.
\item Dispersion changes in strength during an observation as the zenith angle / airmass changes.
\item Dispersion always occurs along the parallactic angle.
\end{enumerate}

\begin{figure}[h]
\includegraphics[scale=0.77]{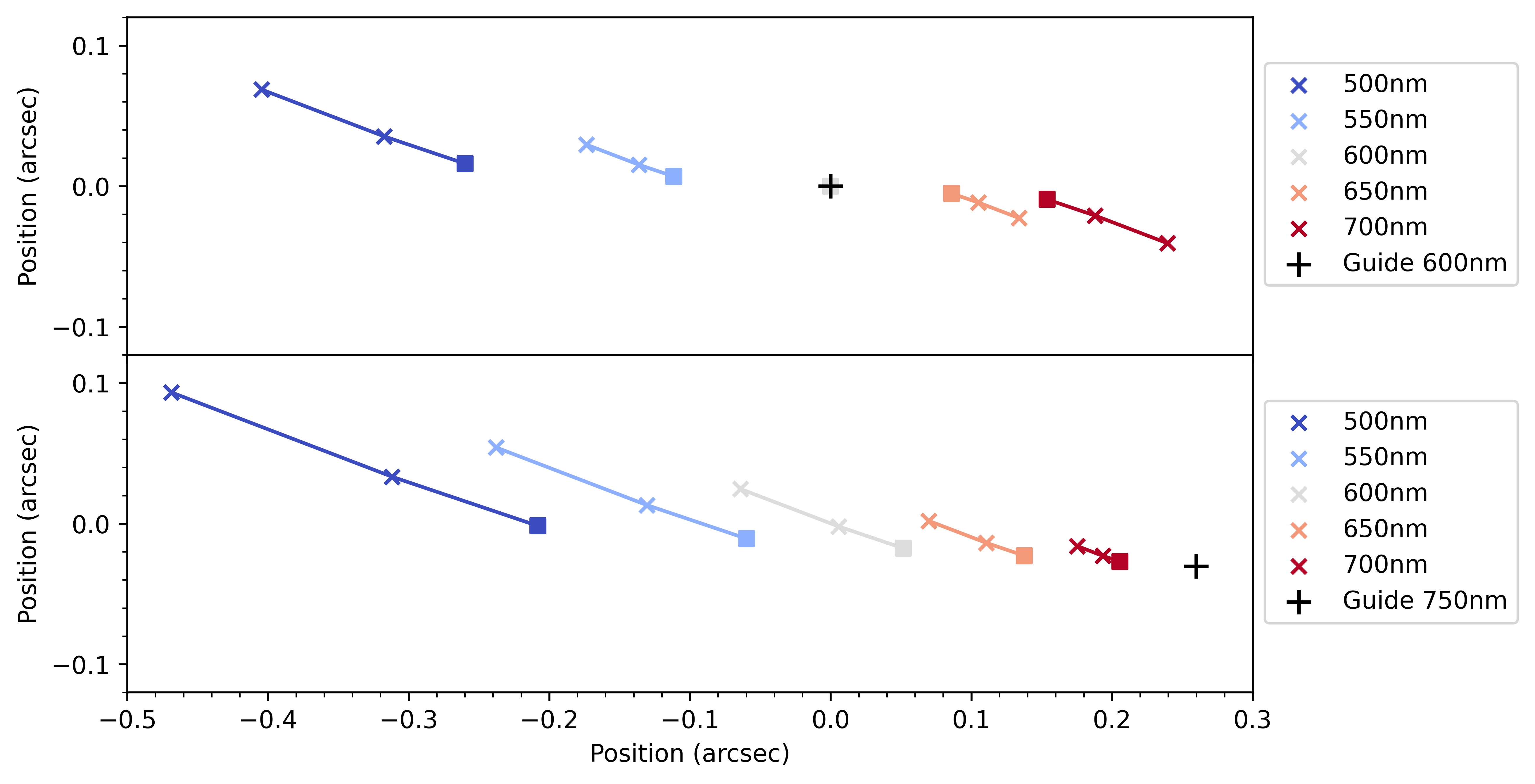}
\centering
\caption{Movement of visible wavelengths on the focal plane during a 1h observation at Paranal due to atmospheric dispersion for two different telescope guiding wavelengths, 600nm and 750nm for the top and bottom figures respectively. This observation is for 3 \textless \ LHA \textless \ 4 and $\delta$ = -30\textdegree. The observation start is denoted by the square, and the crosses represent 30 minute intervals. The magnitude of dispersion is very sensitive to the guiding wavelength, and shifts are larger for bluer wavelengths due to stronger refraction.}
\label{fig:shiftsplot} 
\end{figure}

The movement in part 1 and 2 is calculated with Fillipenko's dispersion model\cite{Fillipenko}, of which a key part is provided in Eq. \ref{eq:dispersion}. The reference wavelength $\lambda_{ref}$ is the telescope guiding wavelength, and the refractive index values $n(\lambda)$ are calculated using the environmental conditions (relative humidity, temperature, and pressure) - the shifts are very sensitive to these environmental conditions. The movement in part 3 is described using a calculation of the parallactic angle\cite{Newcomb}:

\begin{equation}
tan(q) = \dfrac{sin(LHA)}{cos(\delta)tan(\phi)-sin(\delta)cos(LHA)}
\end{equation}
where $q$ is the parallactic angle, $LHA$ is the target local hour angle, $\delta$ is the target declination, and $\phi$ is the observer’s latitude. An example of the resulting movement of light on the focal plane is illustrated in Fig. \ref{fig:shiftsplot}. 

\begin{figure}[h]
\includegraphics[scale=0.8]{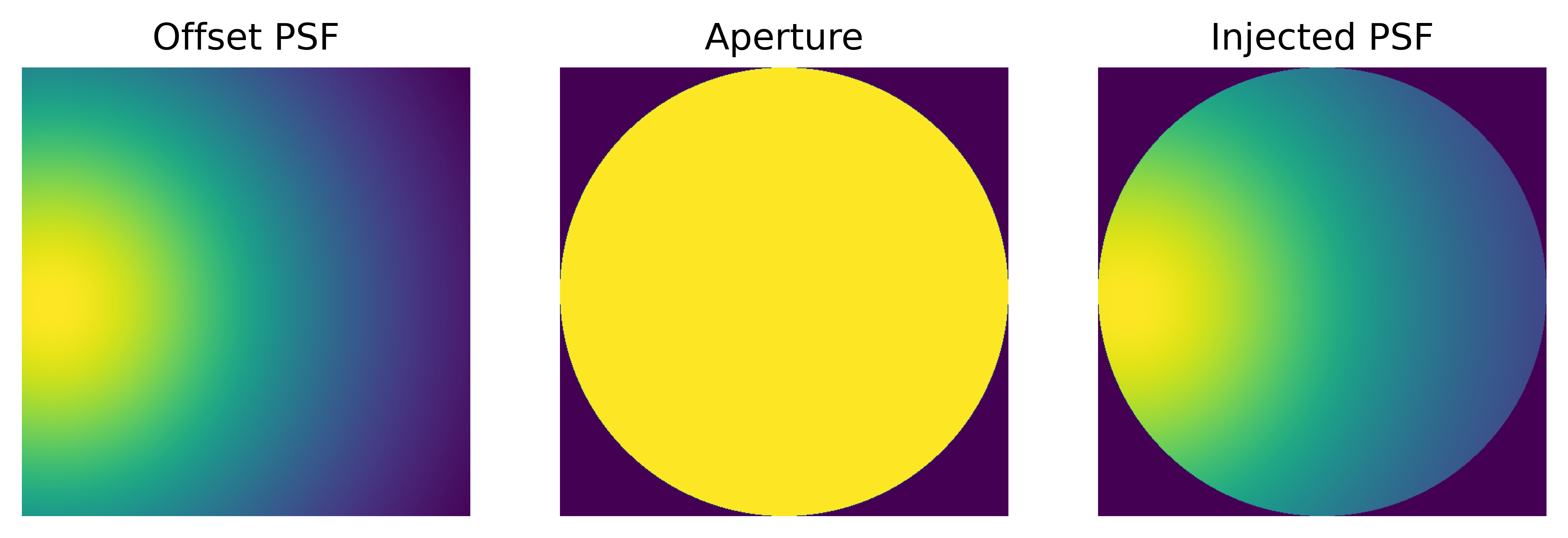}
\centering
\caption{The injection of a monochromatic PSF, $\lambda$=500nm, into a circular aperture, d=0.6", which is shifted due to atmospheric dispersion. 23\% of the PSF's flux is injected into the aperture, relative to 28\% when there is no dispersion, providing a relative injection efficiency, $T_{AD}$, of 82\%}
\label{fig:injection} 
\end{figure}

With the movement of light on the focal plane simulated and an aperture position defined, the displacements of monochromatic PSFs to the aperture are calculated. Every PSF is offset appropriately and injected into the modelled aperture, as seen in Fig. \ref{fig:injection}. The fraction of a PSF's flux that is injected into the aperture is equal to the injected flux divided by the PSF's total integrated flux:

\begin{equation}
T_{raw}(\lambda, X) = \dfrac{\int_{0}^{\infty}\int_{0}^{\infty} I_{PSF,offset}(x,y, \lambda, X) \cdot A(x,y) \cdot dx \cdot dy}{\int_{0}^{\infty}\int_{0}^{\infty} I_{PSF,offset}(x,y,\lambda,X) \cdot dx \cdot dy}
\end{equation}
where $T_{raw}(\lambda,X)$ is the injection efficiency, $I_{PSF,offset}(x,y,\lambda,X)$ is the offset PSF, and $A(x,y)$ is the aperture function. Note that the PSF and the injection efficiency are wavelength ($\lambda$) and airmass ($X$) dependent.

In order to characterise purely the effects of atmospheric dispersion, we need to find the injection efficiency relative to the case where there is no atmospheric dispersion (i.e. when the PSFs are all perfectly centred). This is done by taking the ratio of the offset-PSF and non-offset-PSF injection efficiency:

\begin{equation}
T_{AD}(\lambda, X) = T_{raw}(\lambda, X) /\left(\dfrac{\int_{0}^{\infty}\int_{0}^{\infty} I_{PSF,centred}(x,y, \lambda, X) \cdot A(x,y) \cdot dx \cdot dy}{\int_{0}^{\infty}\int_{0}^{\infty} I_{PSF,centred}(x,y, \lambda, X) \cdot dx \cdot dy}\right)
\end{equation}
where $T_{AD}(\lambda, X)$ is the injection efficiency for the offset-PSF relative to the case where there is no atmospheric dispersion,  and $I_{PSF,centred}(x,y,\lambda,X)$ is the non-offset PSF. By calculating $T_{AD}$ for the observing band wavelengths and observation airmasses (regularly spaced in time) and averaging over time, we produce our injection efficiency curve due to atmospheric dispersion for that instrument and observation.

\section{Results and Discussion}
\label{sec:results}
For a desired instrument configuration and observation details, we can swiftly characterise the atmospheric dispersion impact with Atmosphyre. A comprehensive tutorial of how to use the package can be found on the readthedocs\footnote{https://atmosphyre.readthedocs.io/en/latest/index.html}, and all data and results in this paper were generated using it. What follows is 1) an example test case and general operational recommendations, and 2) a comprehensive analysis of the impact on MOSAIC and benefits of incorporating an ADC. 

\subsection{General Analysis}
\label{sec:General}

\begin{figure}[t]
\includegraphics[scale=0.8]{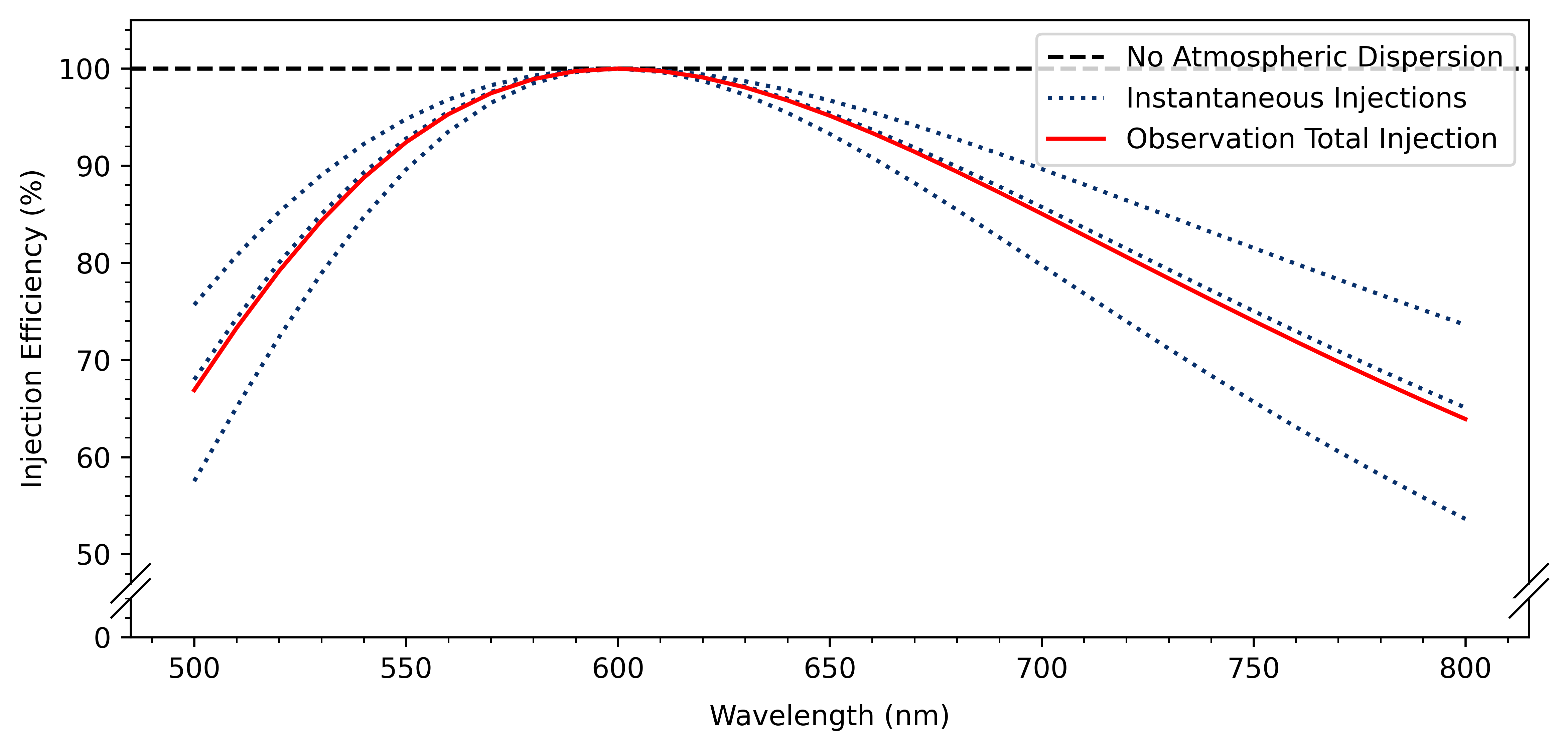}
\centering
\caption{Injection efficiency as a function of wavelength for an observation at Paranal, 3 \textless \ LHA \textless \ 4 and $\delta$ = -30\textdegree. The telescope guides at $\lambda$ = 600nm. The aperture is a circular fibre with a radius of 0.6" and is centred on the guide wavelength (which remains fixed on the focal plane). The observation's total injection efficiency curve (in red) is the time-average of many snapshot injection efficiency curves (3 of which are shown in blue).}
\label{fig:transcurve} 
\end{figure}

In Fig \ref{fig:transcurve}, we present an example injection efficiency curve for a MOS circular aperture operating in the visible. The spectra wing losses and disortions here are around 33\%, and there is a total loss in throughput of 15\%. The differential losses have been minimised and the total throughput maximised with the careful choice of the aperture position and the telescope guide wavelength. This sort of analysis can be used in observation planning (e.g. by setting limits on losses), and to carry-out trade-off studies (e.g. aperture sizes and wavelength bands).

It is difficult to provide an accurate prediction for the optimal aperture position and telescope guide wavelength due to the complex evolution of the dispersion and PSFs through an observation. In order to minimise differential losses and maximise throughput, we find a good rule of thumb for MOS in the visible and NIR is that the optimal guide wavelength is bluer than the observing band's midpoint. The specific value depends on the instrument and observation, primarily the band width and wavelengths. It is usually in the range of 25-45\% of the way through an observing band - in the case of Fig. \ref{fig:transcurve}, the optimal value is the wavelength 33\% of the way through the band. This wavelength remains fixed on the focal plane, and the aperture should be positioned on this point. The optimal wavelength is bluer than the mid-point as shorter wavelengths experience stronger refraction, so this keeps the red and blue wavelengths at comparable offsets. We find that the optimal guide wavelength does not change significantly with a target object's local hour angle and declination, and can be assumed constant for reasonable observations. Atmosphyre includes a feature to determine this optimal guide and fibre position by minimising flux losses and maximising throughput.

If the telescope guide wavelength is fixed at a non-optimal value, then the aperture should be positioned where the optimal wavelength lies on the focal plane at the observation midpoint. During reasonable observations, this is found to keep the flux losses similar to when you are free to choose the telescope guide wavelength. Whether the fixed telescope guide is redder or bluer than the optimal does not matter.

\subsection{MOSAIC ADC Trade-off Study}
To operate on the 39m ELT in the 2030s, MOSAIC will carry out multi-object spectroscopy on the faintest of targets, including distant stars and galaxies throughout cosmic history which cannot be observed with current generation MOS\cite{Hammer}. As part of MOSAIC's development, we used the simulations described in this paper to determine the impacts of atmospheric dispersion on the instrument's baseline design at the time.

\label{sec:MOSAIC}
\subsubsection{Parallel Observations}

\begin{figure}[t]
\includegraphics[scale=0.8]{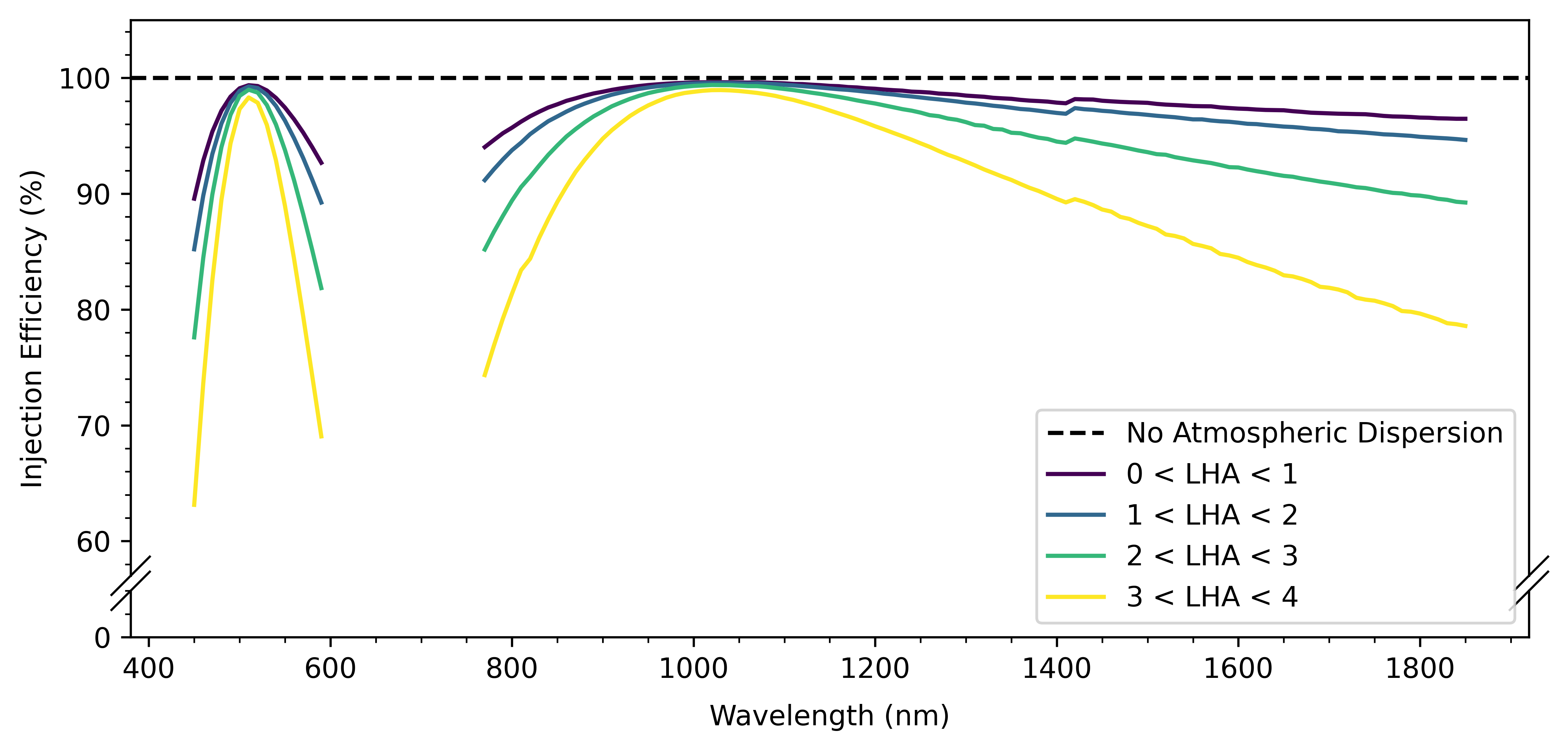}
\centering
\caption{Injection efficiency as a function of wavelength for a parallel-mode MOSAIC observation at $\delta$ = -50\textdegree. VIS-G and NIR are defined as being 450-591nm and 770-1857nm respectively. The guide wavelength is 690nm, and the visible and NIR apertures have been centred on the location of 510nm and 1040nm at the observation midpoint. The PSFs used are from ground layer adaptive optics simulations (credit: MOSAIC/Durham University), and represent a best-case for the performance of the adaptive optics system (hence these losses are worst-case). Bumps in the curve are due to PSFs only being available for discrete wavelengths and airmasses.}
\label{fig:mosaiccurve} 
\end{figure}

MOSAIC is able to observe separate target objects in the visible and NIR simultaneously in a parallel-observing mode\cite{Hammer}. The result is two relevant injection efficiency curves for an observation, as we can see in Fig. \ref{fig:mosaiccurve}\footnote{The given MOSAIC bands are not representative of the final instrument design}. Flux losses have been minimised through the careful choice of the guiding and aperture positions. There are still losses \textgreater \ 5\% for the wings of both the visible and NIR in all 1h observations at $\delta$ =  -50\textdegree. When $LHA$ \textgreater \ 2h these losses are \textgreater \ 10\%. Due to these strong losses, (and other impacts from observing at large angles away from zenith), observations are taken for $\mid LHA\mid$ \textless \ 2h during normal operations. We can clearly see that despite the visible band being far narrower than the NIR band, it experiences greater losses due to the stronger dispersion at blue wavelengths.

The optimal telescope guide wavelength is clearly not as straightforward in the parallel-mode. The value lies somewhere between the individual band's optimal values, typically towards the blue band. As said in Section \ref{sec:General}, the optimal aperture positions are at the location of a wavelength 25-45\% of the way through the bands at the observation midpoint; in Fig. \ref{fig:mosaiccurve}'s case, the wavelengths are 43\% and 25\% of the way through the visible and NIR bands respectively. It is again found that these optimal positions do not change significantly with a target object's local hour angle and declination, and can be assumed constant for reasonable observations.

\begin{figure}[t]
\includegraphics[scale=0.8]{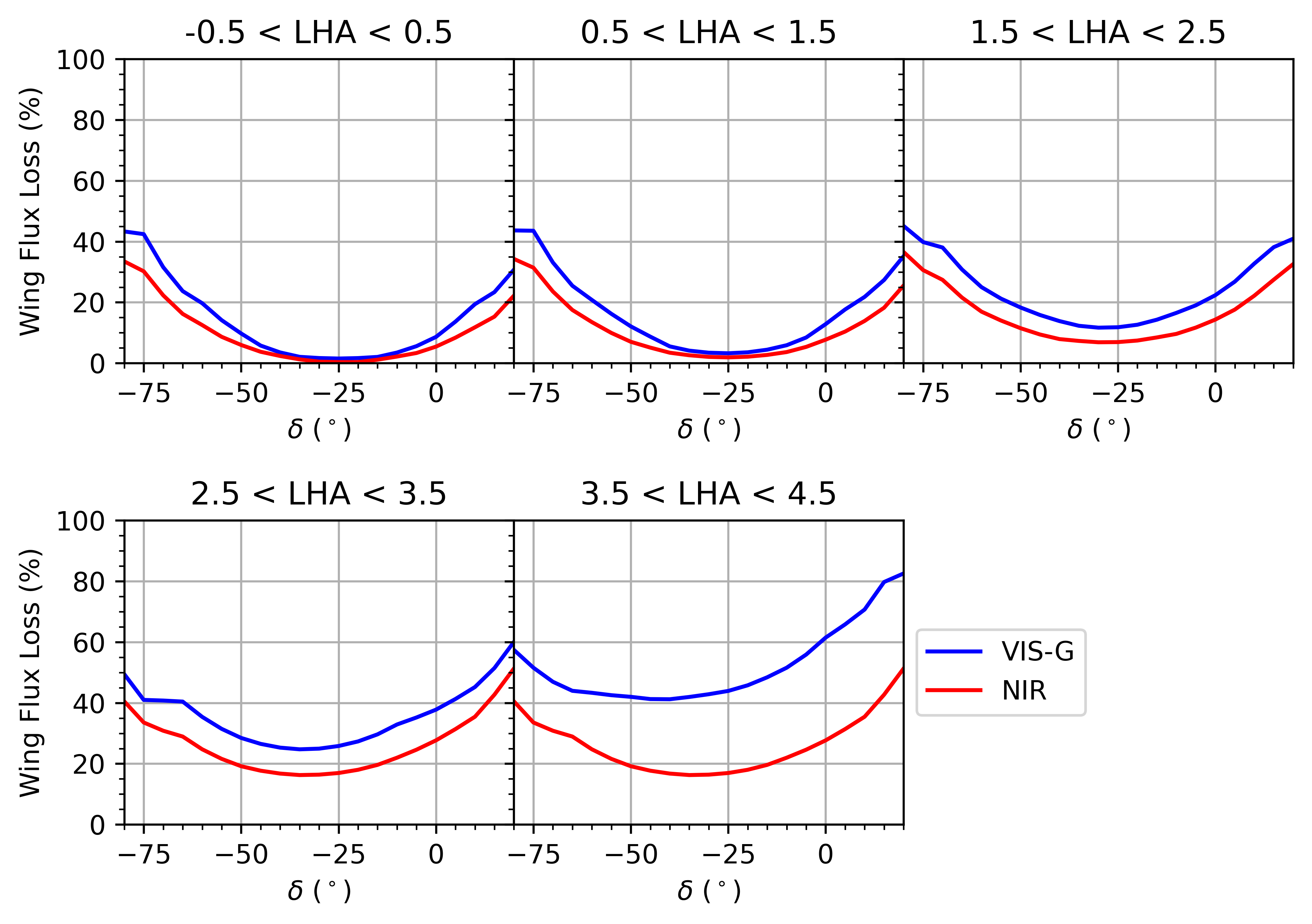}
\centering
\caption{Spectra wing flux losses in the VIS-G and NIR bands as a function of declination for parallel MOSAIC 1h observations in 5 local hour angle ranges. Wing flux loss is defined as 1 - $ T_{AD,min} $ (where $ T_{AD,min} $ is the minimum of a band's injection efficiency for an observation). Bumps in the curve are due to the PSFs only being available for discrete wavelengths and airmasses.}
\label{fig:fluxarraylosses} 
\end{figure}

The spectra wing flux losses for MOSAIC parallel 1h observations at a wide range of declinations and local hour angles can be seen in Fig. \ref{fig:fluxarraylosses}.  Losses have again been minimised. We see that the visible band always has stronger losses than the NIR. The minimum losses are experienced at $\delta$ =  -24.6\textdegree \ (corresponding to the latitude of Paranal) and local hour angles near 0h, as these observations are closest to zenith. Losses dramatically increase at declinations further from this which poses challenges for observing targets near the south pole, such as the Small and Large Magellenic Clouds at $\delta$ = -72.8\textdegree \ and -69.8\textdegree \ respectively. Flux losses \textgreater \ 10\% are unavoidable for for 1h observations with a) $LHA$ \textgreater \ 2.5h, or b) $\delta$ \textless \ -60\textdegree \ and \textgreater \ 10\textdegree \ (corresponding to observations approximately \textgreater  \ 35\textdegree  \ from zenith, or at an airmass \textgreater  \ 1.2).

\subsubsection{Atmospheric Dispersion Corrector Impact}

An atmospheric dispersion corrector (ADC) corrects for the atmospheric dispersion effect by recombining the light of a chromatically dispersed object. Although there will be some residuals, the majority of the wavelength dependent loss is removed, producing a near-flat injection efficiency curve. This results in the removal of most of the spectral distortion, vital for many MOSAIC science cases.

An important metric for MOS is survey speed, which is inversely proportional to the observation time required for each target in the survey to reach a minimum signal-to-noise. It can be derived for a given observation that the survey speed with an ADC, relative to without an ADC, is proportional to the ADC's throughput over the observation's minimum flux injection efficiency, $SS \propto T_{ADC}/T_{AD,min}$. It was found for MOSAIC observations with an ADC at small local hour angles and declinations near -24.6\textdegree \ (corresponding to near zenith / low airmass), there is a slight loss in survey speed. As you move to larger local hour angles and different declinations you start to see an increase in survey speed, with particularly big gains for all visible observations where targets are at $\delta$ \ \textless \ -60\textdegree \ or \textgreater \ 10\textdegree. 

Another important benefit of an ADC is that it unlocks the ability to observe in far wider visible wavelength bands; originally in the baseline design, the visible MOSAIC bands had to be narrow (70 to 190nm wide) in order to limit the effects of atmospheric dispersion. With an ADC, bands spanning much larger wavelength ranges can be chosen, allowing the entirety of the visible to potentially be observed in fewer exposures.

As a result of this characterisation of the atmospheric dispersion impact on MOSAIC, the described benefits, and the consortium's study on ADC designs, it was decided to propose the adoption of ADCs at a positioner level into the baseline design of MOSAIC.

\section{Conclusions}
\label{sec:conclusions}

We've shown that atmospheric dispersion can introduce significant losses and spectral distortions into a spectrograph's observations, being notably stronger for observing bands in the visible. Atmosphyre is a python package able to characterise this impact for a desired instrument and observation, focusing on fibre MOS, and we've shown it can be used for design studies and observation/operations planning.
 
In general we find that the optimal guiding wavelength for observations is bluer than the mid-point of an observing band, between 25-45\% of the way through the band. For a circular aperture, its optimum position is at the location of this wavelength on the focal plane (which remains fixed during an observation). This maximises the throughput and minimises differential losses. If guiding cannot be achieved on this wavelength, then the aperture should be positioned where that optimal wavelength is located on the focal plane at the observation midpoint. The optimal wavelength can be taken as constant for all standard observations, as it does not change significantly for most declinations and local hour angles.

Through characterising the impact of atmospheric dispersion on MOSAIC, we identify flux losses \textgreater \ 10\% are unavoidable with the instrument for 1h observations with a) $LHA$ \textgreater \ 2.5h, or b) $\delta$ \textless \ -60\textdegree \ and \textgreater \ 10\textdegree. The benefits of introducing an ADC to MOSAIC include removing most spectral distortion, increasing the survey speed for many observations away from zenith, and allowing wider visible observing bands to be designed; as a result of this, the adoption of ADCs at the positioner has been proposed for MOSAIC's baseline design.

The next steps for Atmosphyre includes incorporating field differential refraction, an achromatic atmospheric refraction effect - this causes relative position changes between objects in a field-of-view due to differing airmasses\cite{Cuby}. This is particularly important for future wide-field MOS, such as the proposed WST project with a 3 sq. degree field-of-view\cite{WST}, as it cannot be continuously corrected for. Also planned is the addition of aperture positioning error to allow the study of how this effects an instrument's throughput, and the addition of more apertures configurations such as slits.

\appendix    

\section{Atmospheric Contribution to the PSF FWHM}
\label{sec:atmosphericFWHM}

The atmospheric contribution to the image quality is calculated as\cite{EsoETC}:

\begin{equation}
    FWHM_{atm} = s \cdot X^{0.6} \cdot \left(\dfrac{\lambda}{500}\right)^{-0.2} \cdot \gamma
\end{equation}
$s$ is the seeing measured at 500nm at zenith in arcseconds, $x$ is the airmass, lambda is the wavelength in nm, and $\gamma$ is given by:

\begin{equation}
    \gamma = \sqrt{1+F_{Kolb} \cdot 2.183 \cdot \left(\dfrac{r_0}{L_0}\right)^{0.356}}
\end{equation}
$L_0$ is the wave-front outer-scale (calculated to be 46m at Paranal\cite{Ancker}). $r_0$ is the Fried parameter:

\begin{equation}
    r_0 = 0.1 \cdot s^{-1} \cdot (\dfrac{\lambda}{500})^{1.2}\cdot X^{-0.6}
\end{equation}
Here, $s$ is the seeing at 500nm at zenith in arcseconds, $\lambda$ is the wavelength, and $x$ is the airmass.

Finally, $F_{Kolb}$ is the Kolb factor, given by:
\begin{equation}
    F_{Kolb} = \dfrac{1}{1+300 D/L_0}-1
\end{equation}
where $D$ is the telescope diameter in m.

\acknowledgments 
Thanks is given to the MOSAIC consortium and Durham University for help with this work and analysis. Additional thanks is given to ESO.

\bibliography{report} 
\bibliographystyle{spiebib} 

\end{document}